\documentclass[aps,prl,reprint,superscriptaddress,floatfix]{revtex4-2}

\usepackage{amsmath}
\usepackage{amssymb}
\usepackage{bm}
\usepackage{graphicx}
\usepackage{xcolor}
\usepackage[colorlinks=true,
            citecolor=blue,
            linkcolor=blue,
            urlcolor=blue]{hyperref}

\begin{document}

\title{Microscopic Origin of Pressure-Enhanced and Robust Superconductivity in Infinite-Layer La$_{0.8}$Sr$_{0.2}$NiO$_2$}

\author{Jian-Feng Zhang}\email{jianfeng.zhang@hpstar.ac.cn}
\affiliation{ Center for High Pressure Science and Technology Advanced Research, Beijing 100193, China. }
\author{Zhong-Yi Lu}\email{zlu@ruc.edu.cn} 
\affiliation{ School of Physics and Beijing Key Laboratory of Opto-electronic Functional Materials \& Micro-nano Devices, Renmin University of China, Beijing 100872, China }
\affiliation{ Key Laboratory of Quantum State Construction and Manipulation (Ministry of Education), Renmin University of China, Beijing 100872, China }
\author{Tao Xiang}\email{txiang@iphy.ac.cn}
\affiliation{ Institute of Physics, Chinese Academy of Sciences, Beijing 100190, China}
\affiliation{ School of Physical Sciences, University of Chinese Academy of Sciences, Beijing 100049, China}

\date{\today}

\begin{abstract}
Recent transport measurements on freestanding La$_{0.8}$Sr$_{0.2}$NiO$_2$ membranes revealed a broad superconducting dome extending from ambient pressure to $210$~GPa, with an onset transition temperature reaching $74.5$~K near $146$~GPa. Using first-principles calculations, a pressure-dependent two-orbital model, and self-consistent FLEX calculations combined with the linearized Eliashberg equation, we determine how compression modifies the pairing tendency. Pressure increases the kinetic-energy scale, reduces $U_x/t_1$, strengthens interlayer hybridization, and transfers holes from the La/Sr-derived charge reservoir to the correlated Ni sector. Within the present low-energy description, the increasing kinetic scale and the approach to optimal intermediate coupling account for the initial enhancement of pairing, whereas pressure-induced self-doping into the overdoped regime is primarily responsible for its high-pressure suppression. Despite a pronounced three-dimensionalization of the Fermi surface, the pairing-relevant spin susceptibility remains weakly dependent on $q_z$ and peaked near $(\pi,\pi)$. Consequently, the Ni-$d_{x^2-y^2}$-dominated $d$-wave pairing state remains stable over the calculated pressure range. These results provide a unified microscopic interpretation of both the superconducting dome and its unusual robustness under megabar compression.
\end{abstract}

\maketitle

\paragraph{Introduction.---}
Pressure is often regarded as a clean tuning parameter because it introduces no chemical disorder~\cite{presstec}. Microscopically, however, compression simultaneously modifies the bandwidth, effective interaction, dimensionality, and charge transfer. The exceptionally broad superconducting dome recently observed in freestanding La$_{0.8}$Sr$_{0.2}$NiO$_2$ therefore raises two distinct questions: what drives its nonmonotonic pressure dependence, and why does the pairing state survive megabar compression?

The discovery of superconductivity in infinite-layer nickelates~\cite{IN1,IN2} established a new platform for investigating unconventional pairing beyond the cuprates~\cite{cup1,cup2} and iron-based high-$T_c$ superconductors~\cite{iro1,iro2,iro3}. Freestanding nickelate membranes~\cite{IN3} have recently enabled diamond-anvil-cell experiments that were hindered by the thick substrates of conventional films. Pressure was subsequently found to enhance superconductivity strongly in infinite-layer nickelate films and membranes~\cite{pre1,pre2}. Most recently, transport measurements on freestanding La$_{0.8}$Sr$_{0.2}$NiO$_2$ revealed a continuous superconducting dome extending from ambient pressure to $210$~GPa~\cite{pre3}: the onset transition temperature $T_c^{\mathrm{onset}}$ rises nearly linearly from approximately $16$~K to $74.5$~K at $146$~GPa and then decreases gradually to $57.4$~K at $210$~GPa.

In this Letter, we address these questions by combining density functional theory (DFT)~\cite{dft1,dft2}, maximally localized Wannier functions (MLWFs)~\cite{mlwf}, the constrained random-phase approximation (cRPA)~\cite{respack}, and the fluctuation-exchange approximation (FLEX)~\cite{flex1,flex2} with the linearized Eliashberg equation~\cite{flex3}. We show that the dome emerges from the competition between an increasing kinetic-energy scale, a decreasing effective correlation strength, and pressure-induced hole self-doping from the La/Sr charge reservoir. Enhanced interlayer hybridization strongly reshapes the Fermi surface but has only a secondary effect on pairing. Over the calculated pressure range, pairing remains dominated by the Ni-$d_{x^2-y^2}$ orbital and retains $B_{1g}$ $d$-wave symmetry.

\begin{figure*}[t]
\includegraphics[angle=0,scale=0.6]{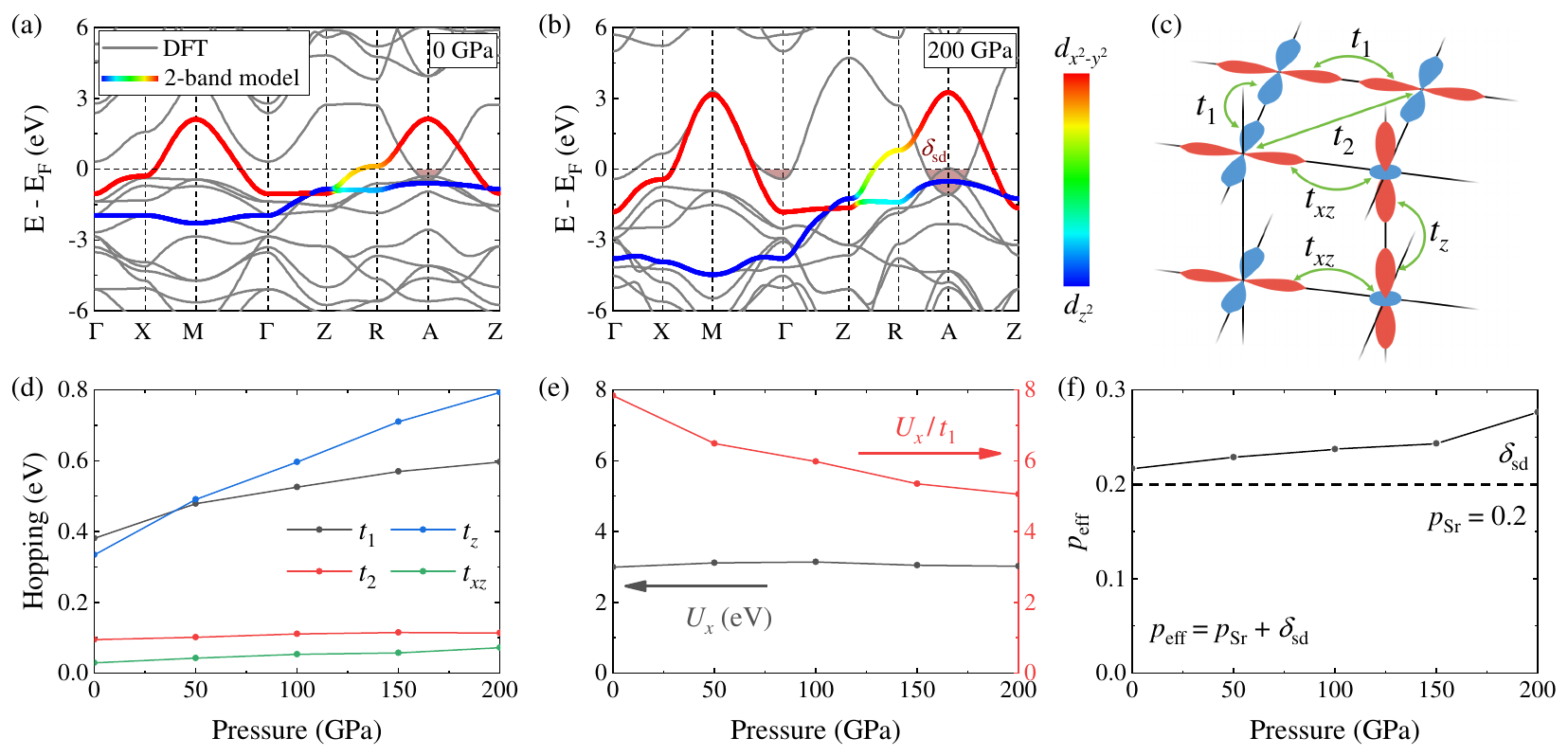}
\caption{Electronic structure and pressure-dependent model parameters of La$_{0.8}$Sr$_{0.2}$NiO$_2$. (a,b) DFT and two-band Wannier dispersions at (a) $0$ and (b) $200$~GPa, with the orbital weights indicated. (c) Leading hopping processes of the effective model. (d) Pressure dependence of the in-plane, interlayer, and interorbital hoppings. (e) Screened interaction $U_x$ and the ratio of $U_x/t_1$. (f) Self-doping $\delta_{\text{sd}}$ and effective hole concentration $p_{\text{eff}}=p_{\text{Sr}}+ \delta_{\text{sd}}$.}
\label{Fig_model}
\end{figure*}

\paragraph{Electronic structure and low-energy model.---}
We first perform DFT structure optimizations and electronic structure calculations for La$_{0.8}$Sr$_{0.2}$NiO$_2$ from ambient pressure to $200$~GPa. Sr substitution is treated within the virtual-crystal approximation. Further computational details are provided in the Supplemental Material.

A two-band tight-binding Hamiltonian containing the Ni-$d_{x^2-y^2}$ and Ni-$d_{z^2}$ orbitals is constructed using MLWFs. As shown in Figs.~\ref{Fig_model}(a) and (b), the Wannier Hamiltonian accurately reproduces the Ni-dominated DFT bands near the Fermi level. The two orbitals play distinct low-energy roles: the $d_{x^2-y^2}$ band forms the principal correlated Fermi surface, whereas the $d_{z^2}$ orbital remains nearly fully occupied throughout the Brillouin zone~\cite{IN-the1,IN-the2}. The La/Sr-derived electron pockets (brown shadows) are retained only in the DFT stage, while their leading influence on the interacting Ni sector is incorporated through a pressure-dependent carrier concentration.

This motivates the following minimal effective Hamiltonian:
\begin{equation}
 \hat{H} = \hat{H}_0 -\sum_{i,\sigma}\mu(\hat{n}_{ix\sigma}+\hat{n}_{iz\sigma})
 + U_x\sum_i \hat{n}_{ix\uparrow}\hat{n}_{ix\downarrow},
\label{eq_Hami}
\end{equation}
where $\hat{H}_0$ is the MLWF tight-binding Hamiltonian, $\mu$ is the chemical potential, and $x$ and $z$ denote the Ni-$d_{x^2-y^2}$ and Ni-$d_{z^2}$ orbitals, respectively. The leading hopping processes are illustrated in Fig.~\ref{Fig_model}(c).

Notably, in Eq.~(\ref{eq_Hami}), only the onsite interaction $U_x$ of the $d_{x^2-y^2}$ orbital is retained in the low-energy pairing vertex. Because the $d_{z^2}$ orbital is nearly fully occupied, its low-energy spin and charge fluctuations are strongly suppressed and provide no efficient pairing channel. Interorbital and Hund interactions are therefore neglected in the present minimal description.

Unlike most spin-fluctuation-driven high-$T_c$ superconductors~\cite{cup1,cup2,iro1,iro2,iro3}, which have predominantly quasi-two-dimensional electronic structures, infinite-layer nickelates possess a sizable interlayer hopping $t_z$ between $d_{z^2}$ orbitals. The interorbital hybridization $t_{xz}$ transfers this strong out-of-plane dispersion to the Fermi surface, producing substantial three-dimensionality even though the active correlated orbital is predominantly $d_{x^2-y^2}$~\cite{IN-the1,IN-the2}. 

Figure~\ref{Fig_model}(d) summarizes the pressure evolution of the leading hopping parameters. The nearest-neighbor in-plane hopping $t_1$ increases from $0.38$~eV at ambient pressure to $0.60$~eV at $200$~GPa, reflecting the enhanced kinetic-energy scale. The interlayer hopping $t_z$ increases even more strongly, from $0.33$ to $0.79$~eV. 
The screened interaction $U_x$ is evaluated using cRPA as implemented in RESPACK~\cite{respack}. In contrast to the hopping parameters, $U_x$ depends only weakly on pressure and remains close to $3.0$~eV [Fig.~\ref{Fig_model}(e)]. Consequently, the dimensionless correlation ratio $U_x/t_1$ decreases from approximately $7.8$ at ambient pressure to $5.1$ at $200$~GPa, driving the system from a strongly correlated regime toward intermediate coupling~\cite{IN-flex1}.
Pressure also expands the La/Sr-derived electron pockets, transferring additional electrons out of the Ni sector. The resulting hole self-doping $\delta_{\text{sd}}$ increases from approximately $0.017$ to $0.076$ holes per Ni [Fig.~\ref{Fig_model}(f)]. Including the nominal Sr concentration $p_{\mathrm{Sr}}=0.20$, the effective hole concentration of the two-band model is
\begin{equation}
 p_{\text{eff}}(P)= p_{\text{Sr}}+\delta_{\text{sd}}(P).
\label{eq_peff}
\end{equation}
In the following calculations, $\mu$ is adjusted at each pressure to reproduce $p_{\text{eff}}(P)$.

\begin{figure*}[t]
\includegraphics[angle=0,scale=0.63]{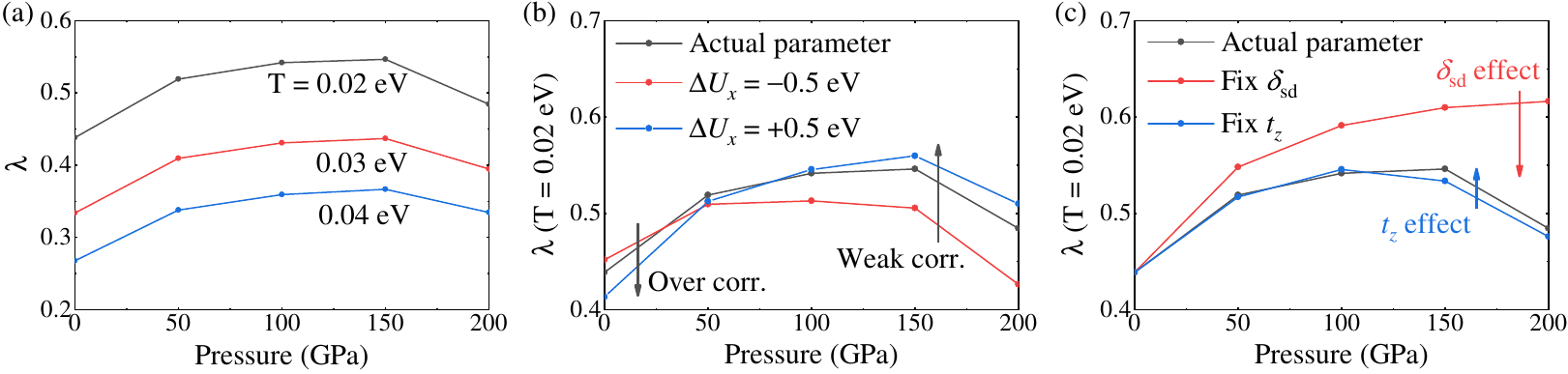}
\caption{Pressure evolution of superconductivity. (a) Leading Eliashberg eigenvalue $\lambda$ at different temperatures. (b) $\lambda$ for the physical interaction $U_x=3.0$~eV and for $U_x$ shifted by $\pm0.5$~eV. (c) Effects of self-doping and interlayer hopping, isolated by fixing $\delta_{\text{sd}}$ or the interlayer hopping parameters at their ambient-pressure values.}
\label{Fig_lam}
\end{figure*}

\paragraph{Pressure evolution of superconductivity.---}
Electronic correlations are treated using self-consistent FLEX~\cite{flex1,flex2,IN-flex1}. The normal self-energy and susceptibilities are calculated from the pressure-dependent Hamiltonian in Eq.~\ref{eq_Hami}, and the resulting spin-singlet pairing interaction is inserted into the linearized Eliashberg equation~\cite{flex3}. The superconducting transition occurs when its leading eigenvalue reaches $\lambda=1$. Here, we use $\lambda$ as a measure of the relative pairing tendency rather than as a quantitative estimate of $T_c$. Because its calculated pressure variation is small, we approximate $U_x(P)$ by a pressure-independent value of $3.0$~eV.

Figure~\ref{Fig_lam}(a) shows the leading Eliashberg eigenvalue at several temperatures. At each temperature, $\lambda$ initially increases rapidly with pressure, develops a weaker slope near $100$~GPa, reaches a maximum around $150$~GPa, and then decreases. This evolution reproduces the qualitative dome-shaped pressure dependence observed experimentally~\cite{pre3}. The leading eigenfunction has $B_{1g}$ $d$-wave symmetry at every pressure investigated~\cite{IN-the3}.

To isolate the microscopic effects of pressure, we selectively vary individual model parameters. First, pressure increases the overall kinetic-energy scale set by $t_1$. Under a uniform rescaling of all energy parameters, the Eliashberg eigenvalue depends only on the reduced temperature $T/t_1$, provided that the dimensionless interaction ratios, band structure, and carrier filling remain unchanged. Thus, at a fixed absolute temperature, an increase in $t_1$ lowers $T/t_1$ and enhances the pairing eigenvalue in the temperature range considered here. The actual pressure evolution is not a uniform rescaling, however, because $U_x/t_1$, the relative interlayer dispersion, and the carrier concentration also change with pressure.

Second, because $U_x$ is nearly pressure independent, the increasing bandwidth substantially reduces $U_x/t_1$. Its influence on the dimensionless pairing strength is nonmonotonic~\cite{IN-flex1}. To demonstrate this effect, we repeat the calculations after shifting $U_x$ by $\pm0.5$~eV, as shown in Fig.~\ref{Fig_lam}(b). For the physical value $U_x=3.0$~eV, the optimal correlation ratio is reached near $50$~GPa, where $U_x/t_1\simeq6.3$, consistent with an earlier dynamical mean-field theory study~\cite{IN-dmft}. At low pressure, the system lies on the overcorrelated side of this optimum. Increasing $t_1$ enhances the antiferromagnetic exchange scale, $J_1\propto t_1^2/U_x$, while also improving quasiparticle coherence. Both effects favor pairing. At higher pressure, the system crosses to the more itinerant side of the optimum, where a further reduction of $U_x/t_1$ weakens the spin-fluctuation pairing vertex. The continuing increase in the absolute kinetic scale nevertheless shifts the maximum of the full superconducting scale to substantially higher pressure.

Third, the pressure-induced expansion of the La/Sr electron pockets increases $\delta_{\text{sd}}$ from $0.017$ to $0.076$, corresponding to an increase in $p_{\text{eff}}$ from $0.217$ to $0.276$. The correlated Ni band is thereby driven progressively into the overdoped regime. When $\delta_{\text{sd}}$ is artificially fixed at its ambient-pressure value, most of the high-pressure reduction of $\lambda$ disappears [Fig.~\ref{Fig_lam}(c)]. Within the present low-energy description, pressure-induced self-doping is therefore the dominant source of the high-pressure suppression of pairing. This result further suggests that reducing the nominal Sr concentration may provide a route toward higher $T_c$. Although such samples may have a lower $T_c$ at ambient pressure, pressure-enhanced self-doping could tune the Ni band toward optimal filling rather than into the overdoped regime, potentially increasing the maximum high-pressure $T_c$~\cite{IN-dmft}.

Finally, pressure strongly enhances the interlayer dispersion: $t_z/t_1$ increases from $0.88$ to $1.33$. Nevertheless, fixing the interlayer hopping parameters at their ambient-pressure values changes $\lambda$ only modestly [Fig.~\ref{Fig_lam}(c)]. The effect becomes discernible only at high pressure and determines neither the initial enhancement nor the downturn of superconductivity. Thus, the pronounced three-dimensionalization of the Fermi surface has only a secondary influence on pairing in the in-plane correlated $d_{x^2-y^2}$ orbital.

The superconducting dome therefore results from the competition among three dominant trends. At low pressure, the increasing kinetic scale and the approach to optimal intermediate coupling enhance pairing, while self-doping remains weak. At intermediate pressure, the further reduction of $U_x/t_1$ begins to oppose the growing kinetic scale, producing a weaker pressure dependence. Above approximately $150$~GPa, accumulated hole self-doping drives the Ni band further into the overdoped regime and generates the descending side of the dome.

\begin{figure*}[t]
\includegraphics[angle=0,scale=0.57]{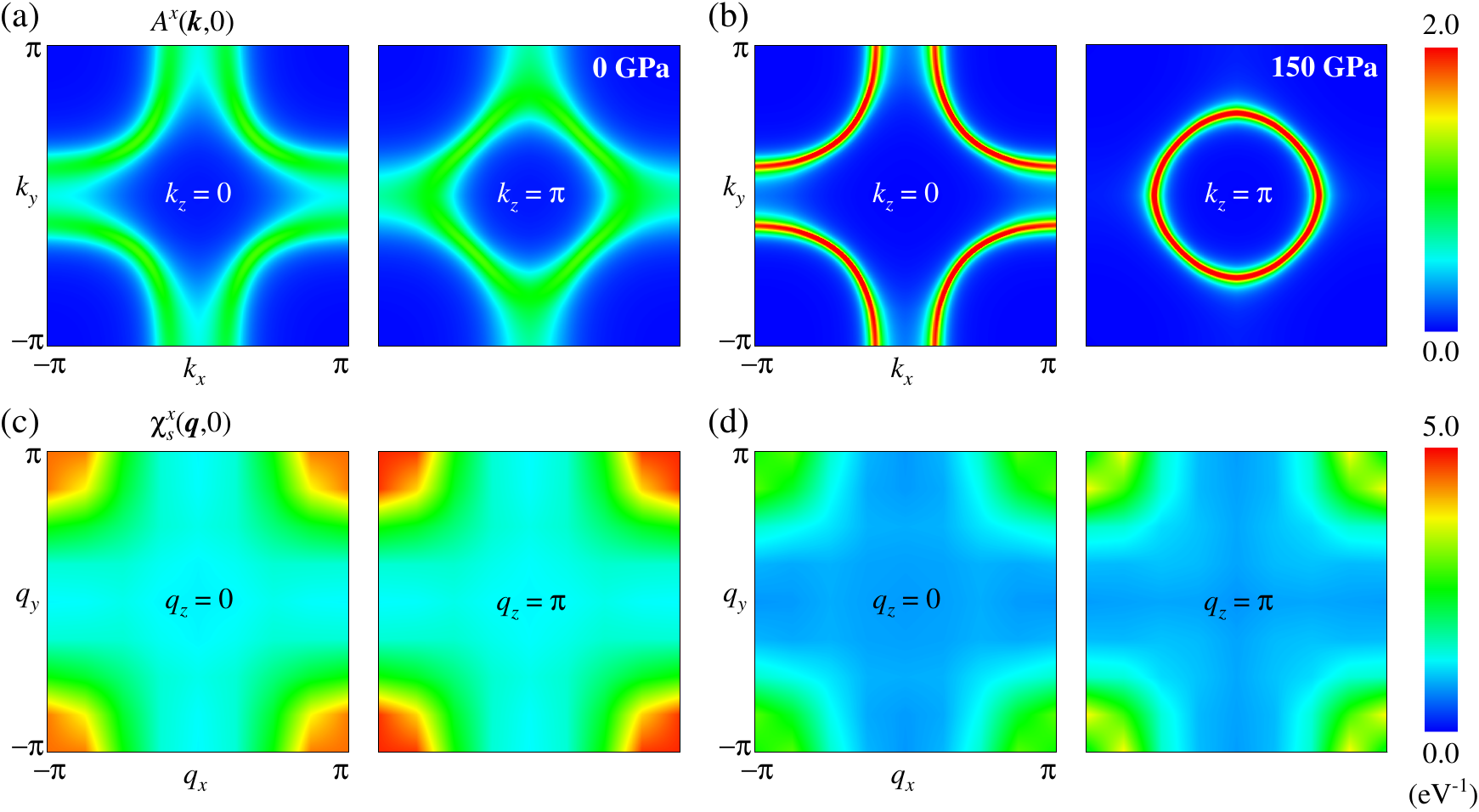}
\caption{Ni-$d_{x^2-y^2}$-resolved electronic and magnetic responses at ambient and optimal pressures. (a,b) Zero-energy spectral function $A^x(\mathbf{k},0)$ on the $k_z=0$ and $\pi$ planes at (a) $0$ and (b) $150$~GPa. (c,d) Corresponding static intra-orbital spin susceptibility $\chi_s^x(\mathbf{q},0)$ on the $q_z=0$ and $\pi$ planes. Pressure sharpens the Fermi-surface spectra and enhances their three-dimensionality, while the susceptibility remains peaked near $(\pi,\pi)$ with weak $q_z$ dependence.}
\label{Fig_spec}
\end{figure*}

\paragraph{Pressure robustness.---}
The robustness of superconductivity in La$_{0.8}$Sr$_{0.2}$NiO$_2$ can be traced to the preservation of its minimal low-energy pairing framework under compression. Figure~\ref{Fig_spec} compares the Ni-$d_{x^2-y^2}$-resolved spectral function $A^x(\mathbf{k},0)$ and static spin susceptibility $\chi_s^x(\mathbf{q},0)$ at ambient pressure and $150$~GPa, near the optimal pressure.

As shown in Figs.~\ref{Fig_spec}(a) and (b), pressure substantially modifies the one-particle electronic structure. The reduction of $U_x/t_1$ improves quasiparticle coherence, producing sharper spectral weight around the Fermi surface at $150$~GPa. At the same time, the enhanced difference between the $k_z=0$ and $k_z=\pi$ sections demonstrates a pronounced three-dimensionalization of the Fermi surface.

The magnetic response evolves much less strongly. Figures~\ref{Fig_spec}(c) and (d) show that $\chi_s^x(\mathbf{q},0)$ remains peaked near the in-plane wave vector $(\pi,\pi)$ at both pressures and on both the $q_z=0$ and $q_z=\pi$ planes. These peaks reflect dominant nearest-neighbor antiferromagnetic fluctuations and favor a $B_{1g}$ $d$-wave gap. Moreover, their weak $q_z$ dependence shows that the pairing-relevant spin fluctuations remain predominantly in-plane, despite the strong three-dimensionality of the Fermi surface. This separation explains the modest influence of $t_z$ on the Eliashberg eigenvalue in Fig.~\ref{Fig_lam}(c).

This behavior contrasts with multiband unconventional superconductors, in which pressure can reconstruct the active electronic states or alter competing pairing channels. Cuprates are sensitive to the charge-transfer energy between Cu-$d_{x^2-y^2}$ and O-$p$ states~\cite{cup-the1,cup-the2,cup-the3}; iron-based superconductors contain several active Fe-$3d$ orbitals and competing ordered states~\cite{iro-the1}; and bilayer La$_3$Ni$_2$O$_7$ possesses a four-orbital correlated manifold formed by the $d_{x^2-y^2}$ and $d_{z^2}$ orbitals of two NiO$_2$ layers~\cite{327}.

By contrast, the active correlated sector of La$_{0.8}$Sr$_{0.2}$NiO$_2$ remains dominated by a single Ni-$d_{x^2-y^2}$ orbital, while the $d_{z^2}$ orbital primarily mediates interlayer dispersion. Pressure changes the bandwidth, correlation strength, dimensionality, and carrier concentration but neither removes the $(\pi,\pi)$ antiferromagnetic fluctuations nor changes the leading $B_{1g}$ pairing symmetry over the calculated range. The pressure evolution of superconductivity thus reflects a continuous tuning of parameters within the same effective framework rather than its reconstruction. This restricted low-energy manifold is consistent with the experimentally observed persistence of superconductivity up to $210$~GPa~\cite{pre3}.

\paragraph{Conclusion.---}
We have established a unified microscopic picture of pressure-enhanced and robust superconductivity in La$_{0.8}$Sr$_{0.2}$NiO$_2$. The increasing kinetic scale and the approach to optimal intermediate coupling enhance pairing at low pressure, whereas pressure-induced self-doping into the overdoped regime produces the high-pressure downturn. Although the Fermi surface becomes strongly three-dimensional, the pairing-relevant spin response remains weakly dependent on $q_z$ and peaked near $(\pi,\pi)$, rendering the effect of interlayer hopping secondary. The resulting preservation of the Ni-$d_{x^2-y^2}$-dominated $B_{1g}$ pairing channel provides a natural explanation for the robustness of superconductivity under megabar compression. More broadly, the restricted low-energy manifold makes infinite-layer nickelates a minimal platform for testing theories of unconventional superconductivity.

\begin{acknowledgments}
This work was supported by the Project funded by China Postdoctoral Science Foundation (No. 2022M723355), the Chinese funding sources applied via HPSTAR, the National Natural Science Foundation of China (12488201), and the National Key Research and Development Project of China (2021ZD0301800, 2022YFA1403103).
\end{acknowledgments}

\bibliography{112}

@Article{IN1,
author={Li, Danfeng
and Lee, Kyuho
and Wang, Bai Yang
and Osada, Motoki
and Crossley, Samuel
and Lee, Hye Ryoung
and Cui, Yi
and Hikita, Yasuyuki
and Hwang, Harold Y.},
title={Superconductivity in an infinite-layer nickelate},
journal={Nature},
year={2019},
month={Aug},
day={01},
volume={572},
number={7771},
pages={624-627},
issn={1476-4687},
doi={10.1038/s41586-019-1496-5},
url={https://doi.org/10.1038/s41586-019-1496-5}
}

@article{IN2,
doi = {10.1088/1361-6633/ac5a60},
url = {https://doi.org/10.1088/1361-6633/ac5a60},
year = {2022},
month = {mar},
publisher = {IOP Publishing},
volume = {85},
number = {5},
pages = {052501},
author = {Nomura, Yusuke and Arita, Ryotaro},
title = {Superconductivity in infinite-layer nickelates},
journal = {Reports on Progress in Physics}
}

@article{IN3,
author = {Yan, Shengjun and Mao, Wei and Sun, Wenjie and Li, Yueying and Sun, Haoying and Yang, Jiangfeng and Hao, Bo and Guo, Wei and Nian, Leyan and Gu, Zhengbin and Wang, Peng and Nie, Yuefeng},
title = {Superconductivity in Freestanding Infinite-Layer Nickelate Membranes},
journal = {Advanced Materials},
volume = {36},
number = {31},
pages = {2402916},
doi = {https://doi.org/10.1002/adma.202402916},
url = {https://advanced.onlinelibrary.wiley.com/doi/abs/10.1002/adma.202402916},
year = {2024}
}

@Article{pre1,
author={Wang, N. N.
and Yang, M. W.
and Yang, Z.
and Chen, K. Y.
and Zhang, H.
and Zhang, Q. H.
and Zhu, Z. H.
and Uwatoko, Y.
and Gu, L.
and Dong, X. L.
and Sun, J. P.
and Jin, K. J.
and Cheng, J.-G.},
title={Pressure-induced monotonic enhancement of Tc to over 30{\thinspace}K in superconducting Pr0.82Sr0.18NiO2 thin films},
journal={Nature Communications},
year={2022},
month={Jul},
day={28},
volume={13},
number={1},
pages={4367},
issn={2041-1723},
doi={10.1038/s41467-022-32065-x},
url={https://doi.org/10.1038/s41467-022-32065-x}
}

@misc{pre2,
      title={High-temperature superconductivity in Nd$_{0.85}$Sr$_{0.15}$NiO$_2$ membranes under pressure}, 
      author={Yonghun Lee and Mengnan Wang and Xin Wei and Yijun Yu and Wendy L. Mao and Yu Lin and Harold Y. Hwang},
      year={2026},
      eprint={2604.09525},
      archivePrefix={arXiv},
      primaryClass={cond-mat.supr-con},
      url={https://arxiv.org/abs/2604.09525}, 
}

@misc{pre3,
      title={Enhanced and robust superconductivity in La0.8Sr0.2NiO2 membranes compressed up to 210 GPa}, 
      author={Shu Cai and Yuqing Tian and Shengjun Yan and Jinyu Zhao and Bo Hao and Jianfeng Zhang and Shuaihang Sun and Yang Ding and Qi Wu and Ho-kwang Mao and I. Bozovic and Yuefeng Nie and Liling Sun},
      year={2026},
      eprint={2608.02042},
      archivePrefix={arXiv},
      primaryClass={cond-mat.supr-con},
      url={https://arxiv.org/abs/2608.02042}, 
}

@article{IN-the1,
  title = {Similarities and Differences between ${\mathrm{LaNiO}}_{2}$ and ${\mathrm{CaCuO}}_{2}$ and Implications for Superconductivity},
  author = {Botana, A. S. and Norman, M. R.},
  journal = {Phys. Rev. X},
  volume = {10},
  issue = {1},
  pages = {011024},
  numpages = {6},
  year = {2020},
  month = {Feb},
  publisher = {American Physical Society},
  doi = {10.1103/PhysRevX.10.011024},
  url = {https://link.aps.org/doi/10.1103/PhysRevX.10.011024}
}

@article{IN-the2,
  title = {Formation of a two-dimensional single-component correlated electron system and band engineering in the nickelate superconductor ${\mathrm{NdNiO}}_{2}$},
  author = {Nomura, Yusuke and Hirayama, Motoaki and Tadano, Terumasa and Yoshimoto, Yoshihide and Nakamura, Kazuma and Arita, Ryotaro},
  journal = {Phys. Rev. B},
  volume = {100},
  issue = {20},
  pages = {205138},
  numpages = {11},
  year = {2019},
  month = {Nov},
  publisher = {American Physical Society},
  doi = {10.1103/PhysRevB.100.205138},
  url = {https://link.aps.org/doi/10.1103/PhysRevB.100.205138}
}

@article{IN-the3,
  title = {Model Construction and a Possibility of Cupratelike Pairing in a New ${d}^{9}$ Nickelate Superconductor $(\mathrm{Nd},\mathrm{Sr}){\mathrm{NiO}}_{2}$},
  author = {Sakakibara, Hirofumi and Usui, Hidetomo and Suzuki, Katsuhiro and Kotani, Takao and Aoki, Hideo and Kuroki, Kazuhiko},
  journal = {Phys. Rev. Lett.},
  volume = {125},
  issue = {7},
  pages = {077003},
  numpages = {6},
  year = {2020},
  month = {Aug},
  publisher = {American Physical Society},
  doi = {10.1103/PhysRevLett.125.077003},
  url = {https://link.aps.org/doi/10.1103/PhysRevLett.125.077003}
}

@article{dft1,
  title = {Inhomogeneous Electron Gas},
  author = {Hohenberg, P. and Kohn, W.},
  journal = {Phys. Rev.},
  volume = {136},
  issue = {3B},
  pages = {B864--B871},
  numpages = {0},
  year = {1964},
  month = {Nov},
  publisher = {American Physical Society},
  doi = {10.1103/PhysRev.136.B864},
  url = {https://link.aps.org/doi/10.1103/PhysRev.136.B864}
}

@article{dft2,
  title = {Self-Consistent Equations Including Exchange and Correlation Effects},
  author = {Kohn, W. and Sham, L. J.},
  journal = {Phys. Rev.},
  volume = {140},
  issue = {4A},
  pages = {A1133--A1138},
  numpages = {0},
  year = {1965},
  month = {Nov},
  publisher = {American Physical Society},
  doi = {10.1103/PhysRev.140.A1133},
  url = {https://link.aps.org/doi/10.1103/PhysRev.140.A1133}
}

@article{mlwf,
title = {An updated version of wannier90: A tool for obtaining maximally-localised Wannier functions},
journal = {Comput. Phys. Commun.},
volume = {185},
number = {8},
pages = {2309-2310},
year = {2014},
issn = {0010-4655},
doi = {https://doi.org/10.1016/j.cpc.2014.05.003},
url = {https://www.sciencedirect.com/science/article/pii/S001046551400157X},
author = {Arash A. Mostofi and Jonathan R. Yates and Giovanni Pizzi and Young-Su Lee and Ivo Souza and David Vanderbilt and Nicola Marzari},

}

@article{flex1,
title = {Conserving approximations for strongly fluctuating electron systems. I. Formalism and calculational approach},
journal = {Annals of Physics},
volume = {193},
number = {1},
pages = {206-251},
year = {1989},
issn = {0003-4916},
doi = {https://doi.org/10.1016/0003-4916(89)90359-X},
url = {https://www.sciencedirect.com/science/article/pii/000349168990359X},
author = {N.E Bickers and D.J Scalapino}
}

@article{flex2,
  title = {Strong-coupling theory of superconductivity in a degenerate Hubbard model},
  author = {Takimoto, Tetsuya and Hotta, Takashi and Ueda, Kazuo},
  journal = {Phys. Rev. B},
  volume = {69},
  issue = {10},
  pages = {104504},
  numpages = {9},
  year = {2004},
  month = {Mar},
  publisher = {American Physical Society},
  doi = {10.1103/PhysRevB.69.104504},
  url = {https://link.aps.org/doi/10.1103/PhysRevB.69.104504}
}

@article{flex3,
  title = {A common thread: The pairing interaction for unconventional superconductors},
  author = {Scalapino, D. J.},
  journal = {Rev. Mod. Phys.},
  volume = {84},
  issue = {4},
  pages = {1383--1417},
  numpages = {0},
  year = {2012},
  month = {Oct},
  publisher = {American Physical Society},
  doi = {10.1103/RevModPhys.84.1383},
  url = {https://link.aps.org/doi/10.1103/RevModPhys.84.1383}
}

@Article{327,
author={Sun, Hualei
and Huo, Mengwu
and Hu, Xunwu
and Li, Jingyuan
and Liu, Zengjia
and Han, Yifeng
and Tang, Lingyun
and Mao, Zhongquan
and Yang, Pengtao
and Wang, Bosen
and Cheng, Jinguang
and Yao, Dao-Xin
and Zhang, Guang-Ming
and Wang, Meng},
title={Signatures of superconductivity near 80{\thinspace}K in a nickelate under high pressure},
journal={Nature},
year={2023},
month={Sep},
day={01},
volume={621},
number={7979},
pages={493-498},
issn={1476-4687},
doi={10.1038/s41586-023-06408-7},
url={https://doi.org/10.1038/s41586-023-06408-7}
}

@article{cup1,
  title = {Correlated electrons in high-temperature superconductors},
  author = {Dagotto, Elbio},
  journal = {Rev. Mod. Phys.},
  volume = {66},
  issue = {3},
  pages = {763--840},
  numpages = {0},
  year = {1994},
  month = {Jul},
  publisher = {American Physical Society},
  doi = {10.1103/RevModPhys.66.763},
  url = {https://link.aps.org/doi/10.1103/RevModPhys.66.763}
}

@article{cup2,
  title = {Doping a Mott insulator: Physics of high-temperature superconductivity},
  author = {Lee, Patrick A. and Nagaosa, Naoto and Wen, Xiao-Gang},
  journal = {Rev. Mod. Phys.},
  volume = {78},
  issue = {1},
  pages = {17--85},
  numpages = {0},
  year = {2006},
  month = {Jan},
  publisher = {American Physical Society},
  doi = {10.1103/RevModPhys.78.17},
  url = {https://link.aps.org/doi/10.1103/RevModPhys.78.17}
}

@article{iro1,
author = {Kamihara, Yoichi and Watanabe, Takumi and Hirano, Masahiro and Hosono, Hideo},
title = {Iron-Based Layered Superconductor $\text {LaO}_{1-x}\text{F}_x\text{FeAs}$ ($x$ = 0.05-0.12) with \text{T}$_c$ = 26 \text{K}},
journal = {J. Am. Chem. Soc.},
volume = {130},
number = {11},
pages = {3296-3297},
year = {2008},
doi = {10.1021/ja800073m},
URL = {https://doi.org/10.1021/ja800073m}
}

@article{iro2,
  title = {Arsenic-bridged antiferromagnetic superexchange interactions in LaFeAsO},
  author = {Ma, Fengjie and Lu, Zhong-Yi and Xiang, Tao},
  journal = {Phys. Rev. B},
  volume = {78},
  issue = {22},
  pages = {224517},
  numpages = {6},
  year = {2008},
  month = {Dec},
  publisher = {American Physical Society},
  doi = {10.1103/PhysRevB.78.224517},
  url = {https://link.aps.org/doi/10.1103/PhysRevB.78.224517}
}

@article{iro3,
author = {Fong-Chi Hsu  and Jiu-Yong Luo  and Kuo-Wei Yeh  and Ta-Kun Chen  and Tzu-Wen Huang  and Phillip M. Wu  and Yong-Chi Lee  and Yi-Lin Huang  and Yan-Yi Chu  and Der-Chung Yan  and Maw-Kuen Wu },
title = {Superconductivity in the \text{PbO}-type structure $\alpha$-\text{FeSe}},
journal = {Proc. Natl. Acad. Sci.},
volume = {105},
number = {38},
pages = {14262-14264},
year = {2008},
doi = {10.1073/pnas.0807325105},
URL = {https://www.pnas.org/doi/abs/10.1073/pnas.0807325105},
}

@article{respack,
title = {RESPACK: An ab initio tool for derivation of effective low-energy model of material},
journal = {Computer Physics Communications},
volume = {261},
pages = {107781},
year = {2021},
issn = {0010-4655},
doi = {https://doi.org/10.1016/j.cpc.2020.107781},
url = {https://www.sciencedirect.com/science/article/pii/S001046552030391X},
author = {Kazuma Nakamura and Yoshihide Yoshimoto and Yusuke Nomura and Terumasa Tadano and Mitsuaki Kawamura and Taichi Kosugi and Kazuyoshi Yoshimi and Takahiro Misawa and Yuichi Motoyama}
}

@Article{presstec,
author={Dubrovinsky, Leonid
and Dubrovinskaia, Natalia
and Prakapenka, Vitali B.
and Abakumov, Artem M.},
title={Implementation of micro-ball nanodiamond anvils for high-pressure studies above 6 \text{Mbar}},
journal={Nat. Commun.},
year={2012},
month={Oct},
day={23},
volume={3},
number={1},
pages={1163},
issn={2041-1723},
doi={10.1038/ncomms2160},
url={https://doi.org/10.1038/ncomms2160}
}

@Article{IN-dmft,
author={Di Cataldo, Simone
and Worm, Paul
and Tomczak, Jan M.
and Si, Liang
and Held, Karsten},
title={Unconventional superconductivity without doping in infinite-layer nickelates under pressure},
journal={Nature Communications},
year={2024},
month={May},
day={10},
volume={15},
number={1},
pages={3952},
issn={2041-1723},
doi={10.1038/s41467-024-48169-5},
url={https://doi.org/10.1038/s41467-024-48169-5}
}

@misc{IN-flex1,
      title={Theoretical study of superconductivity in freestanding infinite-layer nickelate membranes under pressure: mitigation of excess correlation enhances $T_c$}, 
      author={Mahiru Seki and Reo Kono and Naotaka Tanaka and Kensei Ushio and Daiki Nakaoka and Masayuki Ochi and Kazuhiko Kuroki and Hirofumi Sakakibara},
      year={2026},
      eprint={2605.24565},
      archivePrefix={arXiv},
      primaryClass={cond-mat.supr-con},
      url={https://arxiv.org/abs/2605.24565}, 
}

@Article{cup-the1,
author={Keimer, B.
and Kivelson, S. A.
and Norman, M. R.
and Uchida, S.
and Zaanen, J.},
title={From quantum matter to high-temperature superconductivity in copper oxides},
journal={Nature},
year={2015},
month={Feb},
day={01},
volume={518},
number={7538},
pages={179-186},
issn={1476-4687},
doi={10.1038/nature14165},
url={https://doi.org/10.1038/nature14165}
}

@article{cup-the2,
  title = {Effective Hamiltonian for the superconducting Cu oxides},
  author = {Zhang, F. C. and Rice, T. M.},
  journal = {Phys. Rev. B},
  volume = {37},
  issue = {7},
  pages = {3759(R)--3761(R)},
  numpages = {0},
  year = {1988},
  month = {Mar},
  publisher = {American Physical Society},
  doi = {10.1103/PhysRevB.37.3759},
  url = {https://link.aps.org/doi/10.1103/PhysRevB.37.3759}
}

@article{cup-the3,
  title = {Intrinsic electron and hole bands in electron-doped cuprate superconductors},
  author = {Xiang, T. and Luo, H. G. and Lu, D. H. and Shen, K. M. and Shen, Z. X.},
  journal = {Phys. Rev. B},
  volume = {79},
  issue = {1},
  pages = {014524},
  numpages = {5},
  year = {2009},
  month = {Jan},
  publisher = {American Physical Society},
  doi = {10.1103/PhysRevB.79.014524},
  url = {https://link.aps.org/doi/10.1103/PhysRevB.79.014524}
}

@Article{iro-the1,
author={Yi, Ming
and Zhang, Yan
and Shen, Zhi-Xun
and Lu, Donghui},
title={Role of the orbital degree of freedom in iron-based superconductors},
journal={npj Quantum Materials},
year={2017},
month={Oct},
day={18},
volume={2},
number={1},
pages={57},
issn={2397-4648},
doi={10.1038/s41535-017-0059-y},
url={https://doi.org/10.1038/s41535-017-0059-y}
}

\end{document}